\documentclass[11pt]{article}
\usepackage{setspace}
\usepackage{amssymb}
\usepackage{amsmath}
\usepackage{amsfonts}

\def\be{\begin{equation}}

\def\ee{\end{equation}}

\def\dd{\partial}

\newcommand\of[1]{\left( #1 \right)}
\newcommand\sqof[1]{\left[ #1 \right]}

\def\bea{\begin{eqnarray}}

\def\eea{\end{eqnarray}}

\def\half{\frac{1}{2}}

\newcommand\eps{\epsilon}

\begin{document}

\singlespace

\begin{flushright} BRX TH-626 \\
CALT 68-2811
\end{flushright}

\vspace*{.3in}

\begin{center}

{\Large\bf The Bel-Robinson tensor for topologically massive gravity}

{\large S.\ Deser}

{\it Physics Department,  Brandeis University, Waltham, MA 02454 and \\
Lauritsen Laboratory, California Institute of Technology, Pasadena, CA 91125 \\
{\tt deser@brandeis.edu}
}

{\large J.\ Franklin}

{\it  Reed College, Portland, OR 97202 \\
{\tt jfrankli@reed.edu}}

\end{center}

\begin{abstract}
We construct, and establish the (covariant) conservation of, a $4$-index ``super-stress tensor" for topologically massive gravity (TMG).  Separately, we discuss its invalidity in quadratic curvature models and suggest a generalization.\end{abstract}

The $4$-index Bel-Robinson tensor $B_{\gamma\mu\nu\rho}$, quadratic in the Riemann tensor and (covariantly) conserved on Einstein shell, has received much scrutiny in its original $D=4$ habitat (see references in [1]). There, B is the nearest thing to a covariant gravitational stress-tensor, for example playing essentially that role in permitting construction of higher ($L>2$) loop local counter-terms in supergravity [2,3]. It also generalizes to $D>4$, at the minor price of losing tracelessness, like its spin $1$ model, the Maxwell stress-tensor.

In this note, we turn to lower $D$, asking whether B survives in $D=3$ and if so, to what question is it the answer--in what theory, if any, is it conserved? Since the hallmark of $D=3$ is the identity of Riemann and Einstein tensors (they are double-duals), it is obvious that B vanishes identically on pure Einstein (i.e., flat space) shell\footnote{Actually, B can already be made trivial on $D=4$ GR shell,
by adding suitable terms [8].}, and becomes the trivial (and removable) constant tensor $\sim (\Lambda^2 g_{\gamma\mu}\,  g_{\nu\rho} +\hbox{symm})$ in cosmological GR [4]. This leaves the dynamical hallmark of $D=3$, TMG [5], and the new quadratic curvature models [6,7], as the other possible beneficiaries. Our main result is that B both survives dimensional reduction and is conserved on TMG shell, in accord with the similar mechanism ensuring the Maxwell tensor's conservation on topologically massive electrodynamics (TME) shell. Separately, a simple argument shows why it does not work for generic quadratic curvature actions.

One obtains B in $D=3$ by inserting the Riemann-Ricci identites (we use de-densitized $\eps^{\mu\nu\alpha}$ throughout) 
\begin{equation*}
R^{\mu\alpha\nu\beta}\equiv (g^{\mu\nu} R^{\alpha\beta} + \hbox{symm}) \equiv  \eps^{\mu\alpha\sigma} \, G_{\sigma\rho} \, \eps^{\nu\beta\rho}
\end{equation*}
 into a $D=4$ B. The resulting combination is:
\begin{equation}
B_{\gamma\mu\nu\rho} = \bar R_{\mu\nu} \, \bar R_{\gamma\rho} + \bar R_{\mu\rho} \, \bar R_{\gamma\nu}
- g_{\mu\gamma} \, \bar R_{\nu\beta} \, \bar R^\beta_{\, \, \rho}\, , \, \, \, \, \, \, \, \bar R_{\mu\nu} \equiv R_{\mu\nu} - 1/4 \, g_{\mu\nu} \, R;
\end{equation}
the Schouten tensor $\bar R$ also defines the Cotton tensor below.  B is manifestly symmetric under $(\gamma\mu, \nu\rho)$ pair interchanges (but not totally symmetric here because that depended on special $D=4$ identities). Clearly, B vanishes identically for $\bar R_{\mu\nu}=0$, and reduces to a constant tensor for the cosmological $\bar R_{\mu\nu} =\Lambda \, g_{\mu\nu}$ extension, a term which may even be removed by suitably adding to the definition of B there. Turning to TMG, its field equation is [5]
\begin{equation}
G^{\mu\nu} = \mu^{-1} \, C^{\mu\nu} \equiv \mu^{-1}\, \eps^{\mu\rho\gamma} \, D_\rho \, \bar R_{\gamma}^{\, \, \, \nu}
\end{equation}
The Cotton tensor $C^{\mu\nu}$ is identically (covariantly) conserved, symmetric and traceless, so tracing (2) implies $R=0$, which simplifies on-shell calculations; $\mu$ is a constant with dimension of mass. [Our results will also apply to cosmologically extended TMG [9], much as they do for cosmological GR.] Our question then is whether B of (1) is conserved by virtue of (2). The reason we expect this is the close analogy between TMG and its vector version, TME. The latter model's abelian version (its non-abelian extension is similar), has (flat space) field equations resembling (2), 
\begin{equation}
\dd_\beta\, F^{\alpha\beta}= \half \, \mu \, \eps^{\alpha\gamma\beta} F_{\gamma\beta} \equiv \mu \ ^*F^\alpha, 
\end{equation}
while the analog of B is the Maxwell stress tensor 
\begin{equation}
T_{\hbox{\tiny M} \, \mu\nu}=F_{\mu}^{\, \, \, \beta}\, F_{\nu\beta} -1/4 \, g_{\mu\nu} \, F_{\alpha\beta}\, F^{\alpha\beta}.
\end{equation}
It is indeed conserved on TME shell, as follows: 
\begin{equation}
\dd_\nu\,  T^{\mu\nu} = F^{\mu\beta} \, \dd_\nu\, F^\nu_{\, \, \, \beta} = \mu \, F^{\mu\beta}  \  ^*F_\beta \equiv \mu \, \eps^{\mu\alpha\beta} \ ^*F_\alpha \ ^*F_\beta \equiv 0.  
\end{equation} 
This success motivates seeking a TMG chain similar to (5), schematically,
\begin{equation}
D \, B \equiv R \, \of{ D R - D R} \equiv R \, \eps \, C = \mu^{-1} \eps \, C\, C \stackrel{?}\equiv 0;
\end{equation}
that is, we are hoping to set up a curl so as to use the algebraic identity $D_\alpha \bar R_{\beta\gamma} - D_\gamma \, \bar R_{\beta\alpha} \equiv \eps_{\mu \alpha\gamma} \, C^{\mu}_{\, \, \, \beta}$ as indicated.  
[There is a major distinction between the two models, however. The Maxwell tensor is also the stress tensor of TME since its Chern-Simons term, being metric-independent, does not contribute. Hence conservation is guaranteed a priori here [5], unlike the very existence, let alone conservation, of a B for TMG.] 
Taking the divergence of (1) and using (2) indeed yields
\begin{equation}
D_\gamma \, B^{\gamma\mu\nu\rho} = \sqof{ D^\gamma\, \bar R^{\mu\nu} - D^\mu \, \bar R^{\nu\gamma}  } \bar R_\gamma^{\, \, \, \rho} +\sqof{ D^\gamma\, \bar R^{\mu\rho}  - D^\mu\, \bar R^{\rho\gamma} } \, \bar R_\gamma^{\, \, \, \nu} =\mu \, \eps^{\sigma \gamma\mu} \, \of{C_{\sigma}^{\, \, \, \nu} \, C_\gamma^{\, \, \, \rho} + C_{\sigma}^{\, \, \, \rho} \, C_\gamma^{\, \, \, \nu}} \equiv 0
\end{equation}
where the identity follows by the symmetry under $(\sigma\gamma)$.
This establishes the nontrivial role of B as
a ``covariant" conserved gravitational tensor for TMG. It may thus find uses here similar to those of the original B in classifying GR solutions.  Whether it is relevant to the quantum extensions of these theories is unclear, since $D=3$ GR is finite [10] and TMG may be [11]. 

The other gravitational model of special interest in $D=3$ is the ``new quadratic curvature" theory. 
Its $L= a\, R + b\, \bar R^2$, or even its pure $\bar R^2$ variant, does not conserve B. The reason is obvious and applies as well to all quadratic curvature actions in $D=4$. The divergence of (any) B behaves as $R\, DR$, while the $R^2$ field equations read $DD\, R +R\, R=0$, hence they do not tell us anything about $DR$. So unless $R\, DR$ vanishes for algebraic reasons, and it does not, there is no hope already at linearized, $DD\, R$, level, quite apart from the $RR$ terms. A clear example is the $\bar R^2$ field equation itself,
\begin{equation}
\Box \, \bar R_{\mu\nu} + \of{ g_{\mu\nu} \, \Box - \frac{3}{8} \, D_\mu \, D_\nu} \, R + \of{2\, \bar R_{\mu\alpha} \, \bar R^\alpha_{\, \, \, \nu} - g_{\mu\nu} \, \bar R^{\alpha\beta} \, \bar R_{\alpha\beta}  } = 0.
\end{equation}
B-nonconservation also makes physical sense: one would expect the correct candidate (if any) to have the form $B'=DR \, DR$ to reflect the extra derivatives in $R^2$ actions.

In summary, we have obtained a conserved Bel-Robinson tensor for $D=3$ TMG, despite TMG's third derivative order.  It is, gratifyingly, the reduction of one originally defined for $D=4$ GR, and fits nicely with the Maxwell stress tensor's conservation in TME.  We also noted the unsuitability of B as a conserved tensor in quadratic curvature models, suggesting instead that a modified $B' \sim DR \, DR$ might succeed.  

SD acknowledges support from NSF PHY 07-57190 and DOE DE-FG02-164 92ER40701 grants.

\end{document}